# BES with FEM: Building Energy Simulation using Finite Element Methods

A.W.M. (Jos) van Schijndel
Eindhoven University of Technology
P.O. Box 513; 5600 MB Eindhoven; Netherlands, A.W.M.v.Schijndel@tue.nl

**Abstract:** An overall objective of energy efficiency in the built environment is to improve building and systems performances in terms of durability, comfort and economics. In order to predict, improve and meet a certain set of performance requirements related to the indoor climate of buildings and the associated energy demand, building energy simulation (BES) tools are indispensable. Due to the rapid development of FEM software and the Multiphysics approaches, it should possible to build and simulate full 3D models of buildings regarding the energy demand. The paper presents a methodology for performing building energy simulation with Comsol. The method was applied to an international test box experiment. The results showed an almost perfect agreement between the used BES model and Comsol. These preliminary results confirm the great opportunities to use FEM related software for building energy performance simulation.

## 1. Introduction

An overall objective of energy efficiency in the built environment is to improve building and systems performances in terms of durability, comfort and economics. In order to predict, improve and meet a certain set of performance requirements related to the indoor climate of buildings and the associated energy demand, building energy simulation (BES) tools are indispensable. Due to the rapid development of FEM software and the Multiphysics approaches, it should possible to build and simulate full 3D models of buildings regarding the energy demand. Because BES and FEM have quite different approaches the methodology of this research is very important: Step 1, start with a simple reference case where both BES and FEM tools provide identical results. Step 2, add complexity and simulate the effects with both tools. Step 3, compare and evaluate the results. The paper is organized as follows: Section 2 provides an introduction to BES modeling.

Section 3 presents the methodology and preliminary results on BES using Comsol. The paper ends with the conclusions.

## 2. BES: Building Energy Simulation

The website the of U.S. Department of energy (Energy.gov 2012) provides information on about 300 building software tools from over 40 countries for evaluating energy efficiency, renewable energy, and sustainability in buildings. Commonly used within these tools are: Zonal approaches of the volumes, assuming uniform temperatures in each zone and 1D modeling of the walls.

The earliest developments of HAMBase originate from 1988, by prof. Martin H. de Wit. Since 1995, this thermal-hygric model, has become available in MatLab. A short summary of the HAMBase model is presented below, further details can be found in (HAMLab 2012). The HAMBase model uses an integrated sphere approach. It reduces the radiant temperatures to only one node. This has the advantage that also complicated geometries can easily be modelled. In figure 1.1 the thermal network is shown.

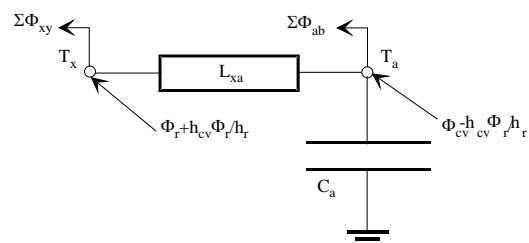

Figure 1 The room model as a thermal network

$T_a$ is the air temperature and $T_x$ is a combination of air and radiant temperature. $T_x$ is needed to calculate transmission heat losses with a combined surface coefficient. $h_r$ and $h_{cv}$ are the surface weighted mean surface heat transfer coefficients for convection and radiation. $\Phi_r$ and $\Phi_{cv}$ are respectively the radiant and convective part of the total heat input consisting of heating or cooling, casual gains and solar gains.* For each heat source a convection factor can be given. For air heating the factor is 1 and for radiators 0.5. The factor for solar radiation depends on the window system and the amount of radiation falling on furniture. $C_a$ is the heat capacity of the air. $L_{xa}$ is a coupling coefficient:

$$L_{xa} = A_t h_{cv}\left(1+\frac{h_{cv}}{h_r}\right) \quad (1)$$

$\sum \Phi_{ab}$ is the heat loss by air entering the zone with an air temperature $T_b$. $A_t$ is the total area. In case of ventilation $T_b$ is the outdoor air temperature. $\sum \Phi_{xy}$ is transmission heat loss through the envelope part $y$. For external envelope parts $T_y$ is the sol-air temperature for the particular construction including the effect of atmospheric radiation.

The thermal properties of the wall and the surface coefficients are considered as constants, so the system of equations is linear. For this system the heat flow entering the room can be seen as a superposition of two heat flows: one resulting from $T_y$ with $T_x=0$ and one from $T_x$ with $T_y=0$. The next equations are valid in the frequency domain:

$$\Phi_x = -\Phi_{xx} + \Phi_{yx} = -Y_x T_x + Y_{xy}(T_y - T_x)$$
$$\Phi_y = \Phi_{yy} + \Phi_{yx} = Y_y T_y + Y_{xy}(T_y - T_x) \quad (2)$$

The heat flow ($\Phi_{yx}$) caused by the temperature difference $\Delta T_{yx}$ is modelled with a fixed time step (1 hour) and response factors. For $t = t_n$:

$$\Phi_{yx}(t_n) = L_{yx}\Delta T_{yx}(t_n) + \Delta\Phi_{yx}(t_n)$$
$$\Delta\Phi_{yx}(t_n) = a_1\Delta T_{yx}(t_{n-1}) + a_2 \Delta T_{yx}(t_{n-2}) + b_1\Delta\Phi_{yx}(t_{n-1}) + b_2\Delta\Phi_{yx}(t_{n-2}) \quad (3)$$

The next equation for the U-value of the wall is valid:

$$A_{xy}U_{xy} = L_{xy} + (a_1 + a_2)/(1-b_1-b_2) \quad (4)$$

For glazing, thermal mass is neglected:

$$L_{yx} = A_{glazing}U_{glazing} \quad (a_1=a_2=b_1=b_2=0) \quad (5)$$

---

*Please note that the model presented in figure 1 is a result of a delta-star transformation.

The heat flow at the inside of a heavy construction is steadier than in a lightweight construction. In such case $L_{yx}$ will be close to zero. In the model $L_{yx}$ is a conductance (so continuous) and $\Phi_{yx}$ are discrete values to be calculated from previous time steps. For adiabatic envelope parts $\Phi_{yx} = 0$. In the frequency domain, the heat flow $\Phi_{xx}$ from all the envelope parts of a room can be added:

$$\Phi_{xx}(tot) = -T_x \sum Y_x \quad (6)$$

The admittance for a particular frequency can be represented by a network of a thermal resistance ($1/L_x$) and capacitance ($C_x$) because the phase shift of $Y_x$ can never be larger than $\pi/2$. To cover the relevant set of frequencies (period 1 to 24 hours) two parallel branches of such a network are used giving the correct admittance's for cyclic variations with a period of 24 hours and of 1 hour. This means that the heat flow $\Phi_{xx}(tot)$ is modelled with a second order differential equation. For air from outside the room with temperature $T_b$ a loss coefficient $L_v$ is introduced. The model is summarized in figure 2

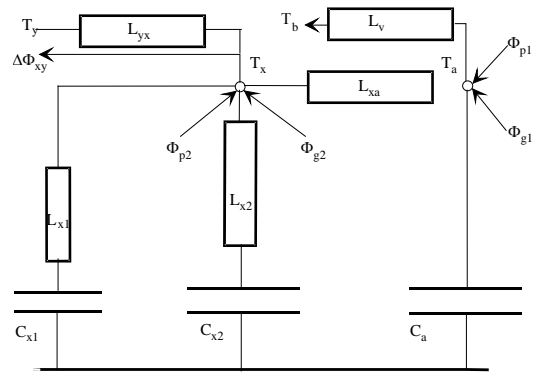

Figure 2 The thermal model for one zone

In a similar way a model for the air humidity is made. Only vapour transport is modelled, the hygroscopic curve is linearized between RH 20% and 80%. The vapour permeability is assumed to be constant. The main differences are: a) there is only one room node (the vapour pressure) and b) the moisture storage in walls and furniture, carpets etc is dependent on the relative humidity and temperature. HAMBase has been validated many times. For the most recent validation study we refer to van Schijndel (2009), where HAMBase was verified using a standard ASHRAE test and validated using the current state-of-the art in the building physics, the IEA Annex 41 test building.

## 3. BES using FEM

### 3.1 Methodology
The methodology was as follows:
Step 1, start with a simple reference case where both BES and FEM tools provide identical results. Step 2, add complexity and simulate the effects with both tools. Step 3, compare and evaluate the results.

### 3.2 Step 1: Reference case
For step 1, a very suitable reference case was found at the current International Energy Agency Annex 58 (2012). It concerns a test box with overall dimension 120x120x120 cm³. Floor, roof and three of the four walls are opaque, one wall contains a window with opening frame. Details of the overall geometry with the exact dimensions can be found in figure 3.

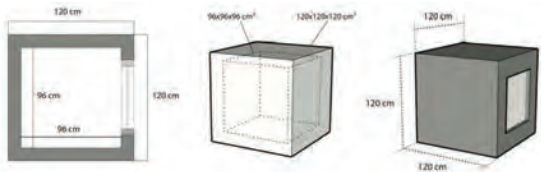

Figure 3. The reference case.

We started to build a 3D model of the opaque test box, heavy weight, air change rate: ACH=0 using Comsol. In order to compare the Comsol 3D FEM model with the HAMBase (HAMLab 2012) lumped model, an equivalent heat conduction of the air is used in Comsol instead of CFD. The distribution in the test box is simulated using Dutch weather data. Figure 4 shows the 3D dynamics snapshots of the isosurfaces. The main challenge now is how to match the high resolution distributed temperature results of Comsol with the lumped temperature results of the BES model. For this reference case (opaque test box, heavy weight, ACH=0) we were able to get a very good match by using a so-called equivalent heat conduction coefficient for the air inside the box in Comsol.

$$k_{eq} = d/R = 1 / 0.34 = 2.9 \qquad (6)$$

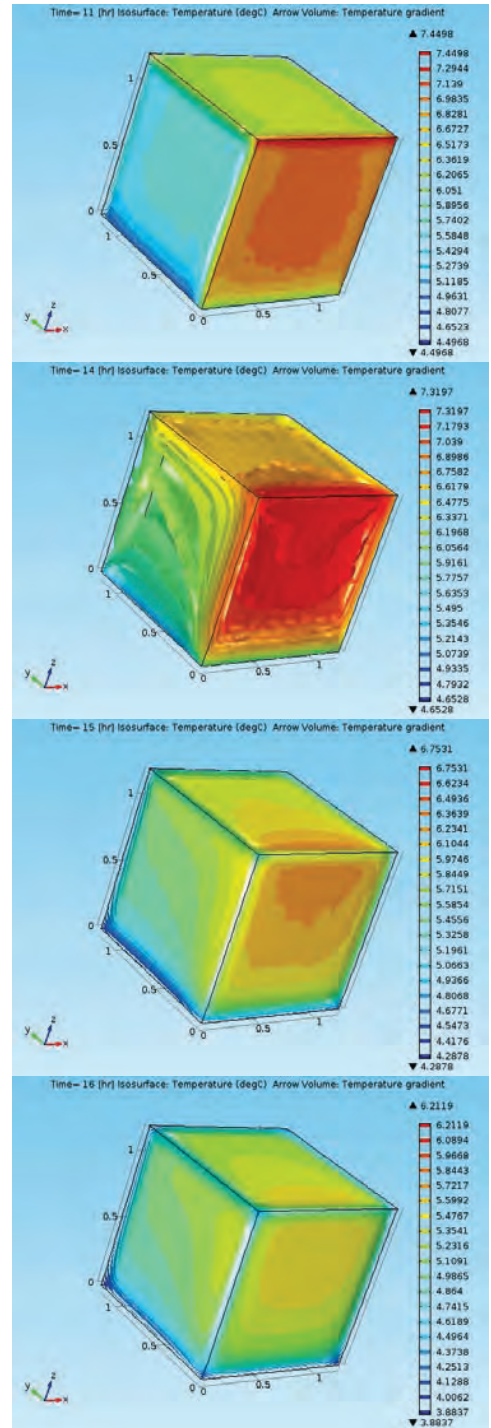

Figure 4 3D dynamics snapshots of the temperature isosurfaces.

Figure 5 shows the comparison of the simulated mean indoor air temperature using Comsol (blue line) and HAMBase (green line) during the first month. The verification result is satisfactory.

**Step 2. Add complexity**

The achievements of the first step i.e. the reference case were quite successful. Therefore we started to add more complexity in the form of solar irradiation. A preliminary detailed result is presented in the appendix. This figure shows the 3D temperature distribution in the test box with a window during the day. Currently we are working on the best way to compare these high resolution distributed temperature results of Comsol with the lumped temperature results of the BES model.

## 4. Conclusions

We conclude that for the reference case Comsol produces identical results as a BES model. The latter is very promising for studying the other required steps of methodology. Furthermore, the presented results are a first step towards more complex 3D FEM simulations including CFD, window, ventilation and radiation. In principle all variants can be simulated in 3D using Comsol, with the notification that the CFD modeling could become quite time consuming.

The 3D modeling allows to virtually place sensors in the test box that produce simulated 'measured' data. This is left over for future research.

These preliminary results confirm the great opportunities to use FEM related software for building energy performance simulation.

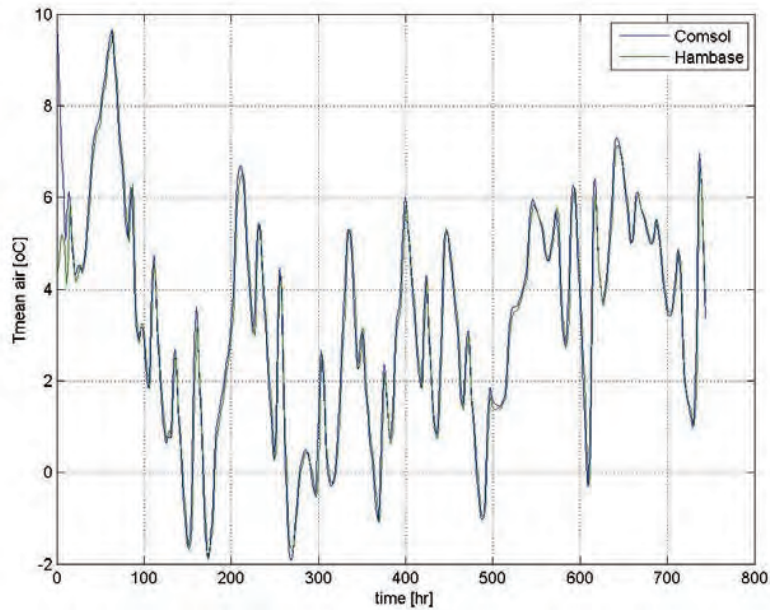

Figure 5. comparison of the simulated mean indoor air temperature using Comsol (blue line) and HAMBase (green line) during the first month.

**Appendix** Preliminary Simulation of the test box with Solar irradiation

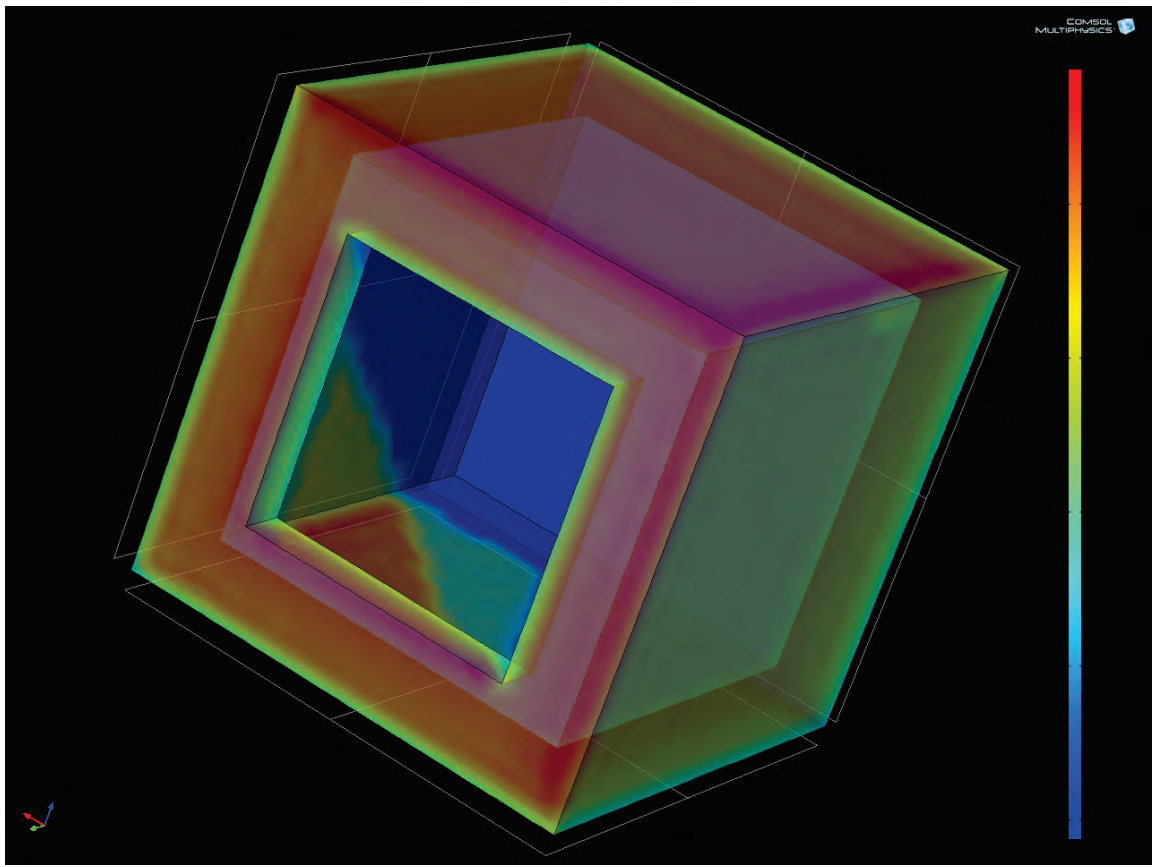